%% file: main.tex
\newif\ifpdflatex    
\pdflatextrue           

\documentclass[fleqn,usenatbib,useAMS,twocolumn]{aastex}
\usepackage{amsmath,esint}
\usepackage{url}
\usepackage{natbib}
\usepackage{soul, CJK}
\usepackage{multirow}
\usepackage{tablefootnote}
\usepackage{float}

\usepackage{bm}
\usepackage{xspace}
\usepackage{color}
\usepackage{enumitem}
\usepackage{booktabs}

\usepackage{graphicx}	
\usepackage{amsmath}	
\usepackage{amssymb}	
\usepackage{bm}		
\usepackage{url}
\usepackage{amsbsy}
\usepackage{xspace}
\usepackage{hyperref}
\usepackage{longtable}
\usepackage[colorinlistoftodos]{todonotes}
\usepackage[flushleft]{threeparttable}
\usepackage{makecell}
\usepackage{csvsimple,booktabs}
\usepackage{filecontents}

\usepackage[T1]{fontenc}
\usepackage{ae,aecompl}

\def\lesssim{\mathrel{\hbox{\rlap{\hbox{\lower5pt\hbox{$\sim$}}}\hbox{$<$}}}}
\def\gtrsim{\mathrel{\hbox{\rlap{\hbox{\lower5pt\hbox{$\sim$}}}\hbox{$>$}}}}

\def\til{\raise.17ex\hbox{$\scriptstyle\mathtt{\sim}$}}

\usepackage{upgreek}                                                  
\usepackage{xspace}                                                       
\input{journals.tex}

\shorttitle{Long Period variable stars from PGIR}
\shortauthors{}

\begin{document}
\title{An Automated Catalog of Long Period Variables using Infrared Lightcurves from Palomar Gattini-IR}
\input{authors_apj.tex}   

\begin{abstract}

Stars in the Asymptotic Giant Branch (AGB) phase, dominated by low to intermediate-mass stars in the late stage of evolution, undergo periodic pulsations, with periods of several hundred days, earning them the name Long Period Variables (LPVs). These stars gradually shed their mass through stellar winds and mass ejections, enveloping themselves in dust. Infrared (IR) surveys can probe these dust-enshrouded phases and uncover populations of LPV stars in the Milky Way. In this paper, we present a catalog of 159,696 Long Period Variables using near-IR lightcurves from the Palomar Gattini - IR (PGIR) survey. PGIR has been surveying the entire accessible northern sky ($\delta > -28^{\circ}$) in the \emph{J}-band at a cadence of 2--3 days since September 2018, and has produced J-band lightcurves for more than 60 million sources. We used a gradient-boosted decision tree classifier trained on a comprehensive feature set extracted from PGIR lightcurves to search for LPVs in this dataset. We developed a parallelized and optimized code to extract features at a rate of \til0.1 seconds per lightcurve. Our model can successfully distinguish LPVs from other stars with a true positive rate and weighted g-mean of 0.95. 73,346 (\til46\%) of the sources in our catalog are new, previously unknown LPVs.\\
\end{abstract}


\date{Last up-dated \today; in original form \today}

\section{Introduction}
\label{sec:intro}
\input{introduction.tex}

\section{Training Sample}
\label{sec:train}
\input{training_sample}

\section{Machine Learning Framework}
\label{sec:ml}
\input{ml_framework}

\section{The Catalog}
\label{sec:cat}
\input{catalog}

\section{Conclusion}
\label{sec:conclusion}
\input{conclusion}

\section*{Acknowledgements}
\input{acknowledgement}

\bibliography{myreferences}
\bibliographystyle{apj}

\label{lastpage}
\end{document}

%% file: journals.tex
\def\apjl{ApJ}%
\def\apjs{ApJS}%
\def\aap{A\&A}%

%% file: authors_apj.tex
\author[0009-0005-8230-030X]{Aswin Suresh}
\affil{Department of Physics, Indian Institute of Technology Bombay, Mumbai 400076, India}

\author[0000-0003-2758-159X]{Viraj Karambelkar}
\affil{Cahill Center for Astrophysics, California Institute of Technology, 1200 E. California Blvd. Pasadena, CA 91125, USA.}

\author[0000-0002-5619-4938]{Mansi M. Kasliwal}
\affil{Cahill Center for Astrophysics, California Institute of Technology, 1200 E. California Blvd. Pasadena, CA 91125, USA.}

\author[0000-0003-1412-2028]{Michael C. B. Ashley}
\affil{School of Physics, University of New South Wales, Sydney NSW 2052, Australia}

\author[0000-0002-8989-0542]{Kishalay De}
\altaffiliation{NASA Einstein Fellow}
\affil{MIT-Kavli Institute for Astrophysics and Space Research, 77 Massachusetts Ave., Cambridge, MA 02139, USA}

\author[0000-0001-9315-8437]{Matthew J. Hankins}
\affil{Arkansas Tech University, Russellville, AR 72801, USA}

\author{Anna M. Moore}
\affil{Research School of Astronomy and Astrophysics, Australian National University, Canberra, ACT 2611, Australia}

\author[0000-0001-9226-4043]{Jamie Soon}
\affil{Research School of Astronomy and Astrophysics, Australian National University, Canberra, ACT 2611, Australia}

\author[0000-0002-4622-796X]{Roberto Soria}
\affil{College of Astronomy and Space Sciences, University of the Chinese Academy of Sciences, Beijing 100049, China}
\affil{INAF-Osservatorio Astrofisico di Torino, Strada Osservatorio 20, I-10025 Pino Torinese, Italy}
\affil{Sydney Institute for Astronomy, School of Physics A28, The University of Sydney, Sydney, NSW 2006, Australia}

\author[0000-0001-9304-6718]{Tony Travouillon}
\affil{Research School of Astronomy and Astrophysics, Australian National University, Canberra, ACT 2611, Australia}

\author{Kayton K. Truong}
\affil{Cahill Center for Astrophysics, California Institute of Technology, 1200 E. California Blvd. Pasadena, CA 91125, USA.}

%% file: introduction.tex

Stars in the final stages of stellar evolution exhibit periodic variations in their brightness on timescales longer than 100 days, and are hence termed as Long Period Variables (LPVs, \citealt{Samus'17}). LPVs are generally categorized into three types: Mira variables, semiregular variables, and OGLE\footnote[1]{Optical Gravitational Lensing Experiment \citep{Udalski92}} small amplitude red giants \citep{Soszynski09, Wray04}. Miras can show oscillation amplitudes larger than ten magnitudes in the optical bands, partly due to absorption of the optical light by nearby molecules. In the infrared, however, the brightness variation is less pronounced. Semiregular variables, which typically display variable amplitudes and periods, and small amplitude red giants are composed primarily of Asymptotic Giant Branch (AGB) stars or Red Giant Branch (RGB) stars and typically show oscillations with amplitudes less than 0.2 mag \citep{Smak66}.

The advent of time domain surveys has resulted in a number of variable star catalogs covering a wide range of optical variability in stellar populations. The Massive Compact Halo Object (MACHO) survey \citep{Alcock2001} found 20000 LPVs in the Large Magellanic Cloud (LMC). \citet{Wood99} used data from 1500 LPVs in the LMC to establish several linear sequences in the log(period) - log(luminosity) space of LPVs. OGLE \citep{Udalski92} identified 91995 LPVs in the Large Magellanic Cloud (LMC) from the OGLE-III catalog of variable stars by examining the log(period) - log(Wesenheit index) relation, finding 1667 Miras in the LMC \citep{Soszynski09}. The Vista Variables in the Vía Láctea (VVV) survey \citep{Dekany11} discovered 129 Miras in the Galactic bulge, along with producing a catalog of 1013 LPVs in the Milky Way using point-spread function photometry data in the {$\text{K}_{\text{s}}$} band from 10 tiles around the Galactic bulge and studied the period-amplitude relationship of Miras in the near-infrared \citep{Nikzat22}. The largest catalog of LPVs to date, consisting of over 1.7 million LPV candidates, of which 392,240 have published periods, was presented in \citet{Lebzelter2023} using 34 months of data from the third Gaia data release. \citet{Karambelkar19} identified 417 extremely luminous long period variables in nearby galaxies from the SPitzer InfraRed Intensive Transients Survey \citep{Kasliwal18} with the \emph{Spitzer Space Telescope} and extended the period-luminosity relationship of AGB stars to longer periods and luminosities. Several catalogs of variable stars consisting of a mixture of short-period (<100 days) stars, such as Cepheids and RR Lyrae, as well as longer-period variables, have been identified (see for e.g. \citealp{chen18}, \citealp{Chen2020}, \citealp{Pojmanski2005}). The wide parameter space of stellar variability and increasingly large datasets from time-domain surveys present the biggest challenge in these studies to identify and characterize variable stars. 

The efficacy of Machine Learning (ML) techniques in identifying and classifying variable stars has been demonstrated in multiple works. For instance, \citet{Masci14} used a random forest classifier trained on lightcurve morphology-based features and Fourier decomposition to distinguish between Algols, RR Lyrae, and W Ursae Majoris among 8273 variables from WISE. \citet{Debosscher07} employ a multivariate Gaussian mixture classifier to classify between 35 classes of variable stars solely using Fourier amplitudes and phases as the feature set. \citet{Sanchez2021} classified 868,371 sources using lightcurves from the Zwicky Transient Facility \citep{Bellm2019} using a multi-level classifier architecture trained on 152 features.  \citet{Boone2019} developed a gradient-boosted decision tree classifier trained on features extracted from Gaussian process augmented multi-band lightcurves to classify between several variable and transient classes ranging from Miras and RR Lyrae to TDEs, SNe\,Ia, and kilonovae.

In this paper, we employ an ML approach to search for Long Period Variables using data from the Palomar Gattini-IR (PGIR) survey taken between September 2018 and July 2021. PGIR’s survey of the northern sky in the Near-Infrared J band, at a cadence of 2 days to a depth of 16 AB mag, has produced lightcurves of more than 60 million stars. We utilize a machine learning framework to identify LPVs in this lightcurve archive. We build a comprehensive training set consisting of LPVs, other erratic and non-periodic variables such as RCrB stars, Young Stellar Objects and non-variables. We train a gradient-boosted decision tree to distinguish between these classes, and use its predictions to generate a catalog of 159,696 LPVs. Of these, 73,346 (45.9\% of the full catalog) are newly identified LPVs. We describe the training sample and feature set for the machine learning classifier in Section \ref{sec:train}. We explain the training procedure, iterative build-up of the training set, and the relative importance of various features in Section \ref{sec:ml}. We analyze the features of LPVs in the PGIR catalog and assess the validity, completeness, and purity of the catalog by comparison with other catalogs in Section \ref{sec:cat}. Section \ref{sec:conclusion} summarizes our conclusions.

%% file: training_sample.tex
The default data processing pipeline of PGIR \citep{De2020} produces light curves for sources detected in multi-epoch images as part of `match files'. Briefly, the reference images of each field is used to construct a master source catalog, and spatially cross-matched to the source catalog for every observation of the field as part of survey operations. The resulting cross-match produces light curves for every source detected in the reference image covering the epochs where the source is detected in the individual epochs. In this paper, we use data from the beginning of PGIR survey operations until 14 July 2021, giving a baseline of \til1400 days for each lightcurve, which is well-suited for characterizing the few hundred day periods of LPVs. 

Constructing a catalog of Long Period Variables requires distinguishing LPVs from non-periodic variable stars (referred to as ``non-LPVs" hereafter) and non-variable stars (``non-vars" hereafter) with high accuracy.  The vast parameter space spanned by these stellar lightcurves makes machine learning a viable option to learn intricate features distinguishing the three classes. The starting point for building the machine learning classifier is assembling a training dataset representative of the various characteristics of LPVs, non-LPVs and non-vars. We build this training set iteratively. We first train a classifier on  a small set of bonafide members of each category. We use this classifier to predict the classes of a larger sample of sources, visually examine the results and use the correctly classified sources to expand the initial training set. The initial training set is described in Section \ref{sec:initial_training} and the final, expanded dataset is described in \ref{sec:expanded_training}. 


\subsection{Initial bonafide training set}
\label{sec:initial_training}
\begin{figure*}
\epsscale{1.2}
\plotone{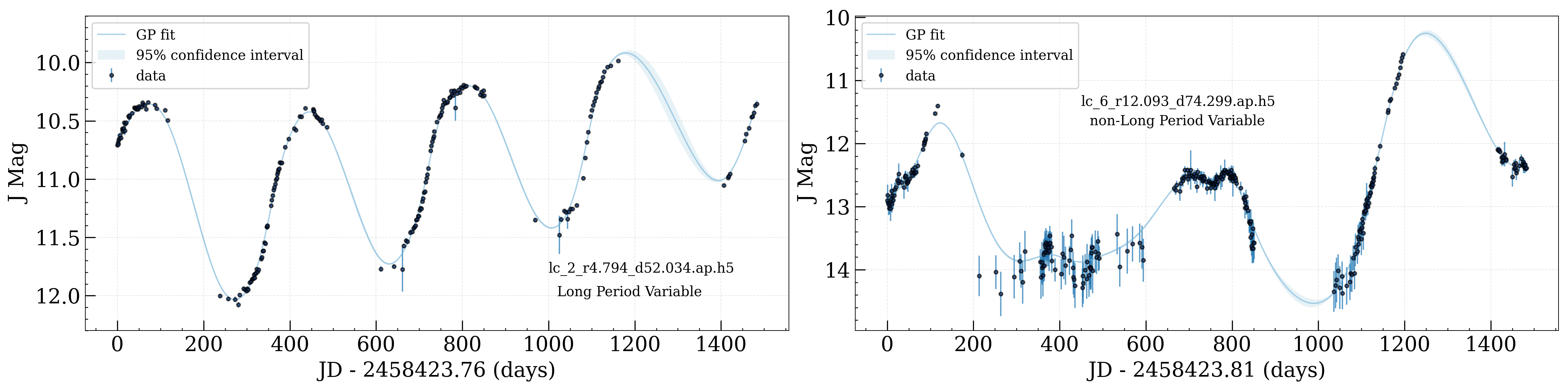}
\caption{Examples of LPVs and non-LPVs from the bonafide set. LPVs display an approximately sinusoidal morphology in their lightcurves, as can be seen in the figure on the left, while non-LPVs show erratic behavior, as the spectroscopically confirmed RCorBor variable shown on the right. The blue line and shaded blue region represent the mean of the Gaussian process regression fit and the 95\% confidence interval respectively.
\label{fig:example_lpv_nonlpv}}
\end{figure*}

We begin with a relatively small training set of 1344 sources, consisting only of LPVs and non-LPVs. These were assembled as part of ongoing searches for the erraticaly varying R Coronae Borealis (R Cor Bor, \citealt{Clayton12}) stars and other erratic, large-amplitude Galactic variables using PGIR (\citealp{Karambelkar2021},  Earley et al. 2024, in prep.).  Our ``LPV" set includes stars identified as LPVs in these studies, primarily based on visual inspection of their lightcurves and Lomb Scargle \citep{Lomb76,Scargle82} periodograms. We also include any additional late M-type stars that were spectroscopically classified as part of these searches. The ``non-LPV" set consists of all known, spectroscopically confirmed R Cor Bor stars that show erratic variations in PGIR data and a smattering of other erratically varying stars such as young stellar objects. 
LPVs constitute the majority sample of this initial training set, with 1265 samples, and non-LPVs form the minority class with 79 samples (5.87\% of the full dataset). Representative examples of the two classes are shown in Figure \ref{fig:example_lpv_nonlpv}. 

This training sample may not be sufficient to cover the breadth of parameters spanned by the two classes. Hence it is used as a baseline dataset to obtain broad classifications. The classifier will inevitably have a high misclassification rate in the first few training iterations before a more balanced dataset enables it to learn the characteristics of both classes. To mitigate this, we built a bigger dataset iteratively by visually examining the predictions of previous classifiers and re-labeling misclassified sources manually (see Section \ref{sec:expanded_training}). 

\subsection{Feature Extraction}
\label{ssec: features}
We use a set of 16 features for initial training iterations. 
We begin by fitting a Gaussian Process (GP) \citep{Williams1995} to model the lightcurves. Gaussian processes have been shown to be an effective method for astronomical lightcurve modeling (see for e.g. \citealp{Boone2019}, \citealp{Qu2021}, \citealp{Foreman2017}). We use the python package \texttt{george} \citep{Ambikasaran2015} to implement the GP fits. The kernel of choice is a Matern 3/2 kernel with a length scale set to the period extracted from the raw lightcurve. A uniform time sampling of 500 points from the first observation to the most recent observation is used to calculate the predicted J-band magnitudes. The minimum and maximum slopes and peak-to-peak amplitudes are calculated from the augmented lightcurve. The goodness of fit of the Gaussian process is quantified by the $\textit{R}^2$ score and is saved as the ``GP score". We calculate the difference between the $90^{th}$ percentile and $50^{th}$ percentile of the J-band magnitudes of each lightcurve as the percentile difference.

Periodicity-based features are extracted using the Lomb-Scargle periodogram, implemented using the \texttt{gatspy} library \citep{VanderPlas2015}. The augmented lightcurve, with J band magnitude, converted to flux, is fit to obtain the three most prominent periods and their associated Lomb-Scargle score. A sinusoid with the best fit period is fit to the lightcurve to calculate the reduced $\chi^2$ of the fit. Additionally, the reduced null $\chi^2$ is calculated as 
\begin{equation}
\chi^2_{red, null} = \sum_i \left(\frac{y_i - \hat{y}}{\sigma_i}\right)^2
\end{equation}
where ${y_i}$ and $\sigma_i$ are the fluxes and errors respectively, and $\hat{y}$ is the error weighted mean of J band flux.

The final periodicity-based features used from these calculations are the Lomb-Scargle score of the best period - LS score, reduced $\chi^2$ and reduced null $\chi^2$, ratio of the best fit period to the maximum period (defined as the observation baseline) - period ratio, ratio of Lomb-Scargle score of the two most prominent periods - Lomb-Scargle score ratio and ratio of reduced $\chi^2$ and reduced null $\chi^2$ - ${\chi^2}$ ratio. Phase-folded lightcurves of LPVs show a clean sinusoidal evolution. The $\chi^2$ value of a sinusoidal fit to the phase folded lightcurve is saved as ``phase $\chi^2$" and is useful to distinguish non-LPVs, non-variables, and erratic variables, which have high phase $\chi^2$ values compared to LPVs.

Another useful measure of variability is given by the Stetson L index \citep{Stetson96}:

\begin{equation}
L = \frac{JK}{0.789}
\end{equation}

where,

\begin{equation}
J = \sum_{n = 1}^{N-1}{\text{sign}(\delta_{n}\delta_{n+1})\sqrt{|\delta_{n}\delta_{n+1}|}}
\end{equation} 

and

\begin{equation}
K = \frac{1/N \sum_{i = 1}^{N}{|\delta_i|}}{\sqrt{{1/N} \sum_{i = 1}^{N}{{\delta_i}^2}}}
\end{equation} 

are the Stetson J and K indices \citep{Stetson96}, with $\delta_i$ being the difference between the magnitude of the $i^{th}$ point and the mean magnitude of the lightcurve, scaled by the error of the $i^{th}$ point.

\begin{figure}[h!]
\epsscale{1.2}
\plotone{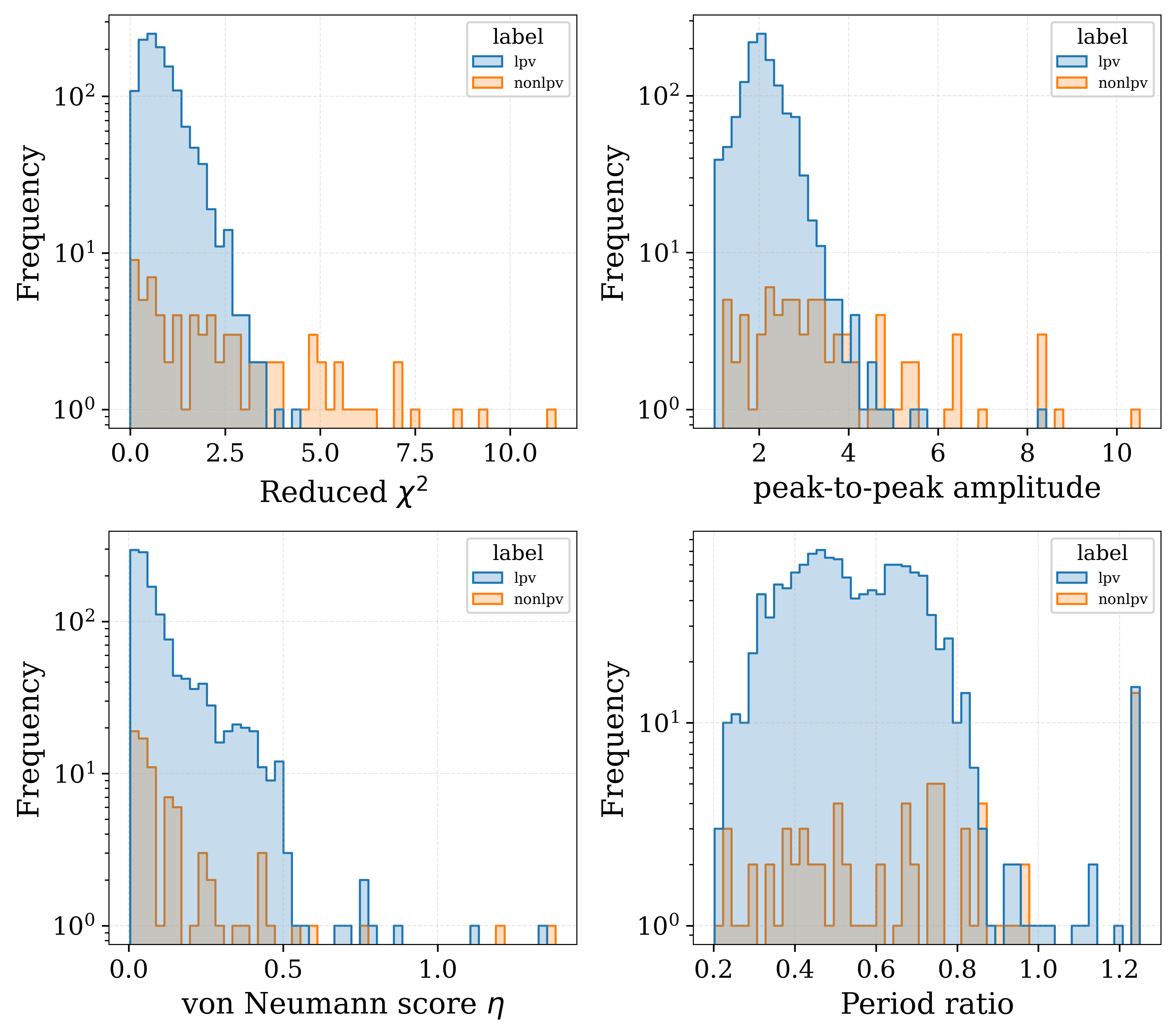}
\caption{Distributions of select features for variables in the bonafide set. Features such as the Reduced $\chi^2$ and the peak-to-peak amplitude can be seen to separate the LPVs and non-LPVs while the period ratio and the von Neumann score occupy similar values for both classes.
\label{fig:gp_features_hist}}
\end{figure}

The final feature extracted from lightcurves is the von Neumann score \citep{Neumann1941} calculated as
\begin{equation}
\eta = \frac{\sum_{n=1}^{N-1} (x_{n+1} - x_n)^2 / (N-1)}{\sigma^2}
\end{equation}.
$\eta$ is a measure of the correlation between successive observations, and positive correlations lead to small values of $\eta$. 

Using a fully parallelized and optimized feature extraction process, our code can calculate all features for each lightcurve in $\sim 0.1$ seconds. The distribution of select features for LPVs and non-LPVs in the bonafide set is shown in Figure \ref{fig:gp_features_hist}


We supplement the features obtained directly from PGIR lightcurves using color information of variable stars observed by other ground and space-based missions. In particular, we query the J, H, and K band colors of variables from the 2-Micron All Sky Survey (2MASS) \citep{Skrutskie06}. The 2MASS All-Sky Catalog of Point Sources \citep{Cutri2003} contains over 300 million objects, including variable stars. We query the infrared photometry of crossmatched sources and calculate the J-H and J-K colors. We note that the J-H and J-K colors are highly correlated in the bonafide set. The Wide-field Infrared Survey Explorer (WISE) \citep{Wright2010} is a space-based infrared observatory surveying the infrared sky at 3.4 (W1), 4.6 (W2), 12 (W3) and 22 (W4) ${\mu}m$ wavelengths. We query the WISE All-sky data release \citep{Cutri2012} to obtain W1-W2, W1-W4, and W3-W4 colors. For cases where no crossmatch is present in WISE or 2MASS, the median color value of the entire dataset is used.

\begin{deluxetable}{ll}
\tablecaption{Summary of features used to train the classifier.
\label{tab:feature_list}}
\tablehead{
    \colhead{Features} &
    \colhead{Description}
}
\startdata
    \texttt{von\_neumann\_score} & {von Neumann score $\eta$} \\
    \texttt{ptp} & {Peak-to-peak amplitude} \\
    \texttt{slope\_min} & {Minimum slope of lightcurve} \\
    \texttt{slope\_max} & {Maximum slope of lightcurve} \\
    \texttt{gp\_score} & {$\textit{R}^2$ of GP fit} \\
    \texttt{per\_diff\_90\_50} & \makecell[l]{Difference between $90^{th}$ and $50^{th}$ \\ percentiles of J-band magnitude} \\
    \texttt{LSscore} & {Lomb-Scargle score of best period} \\
    \texttt{redchi2} & {Reduced $\chi^2$ of sinusoid fit} \\
    \texttt{rednullchi2} & \makecell[l]{Reduced $\chi^2$ assuming constant \\ expected value} \\
    \texttt{chi2ratio} & {Ratio of \texttt{redchi2} and \texttt{rednullchi2}} \\
    \texttt{period\_ratio} & \makecell[l]{Ratio of best period and \\ maximum period} \\
    \texttt{LSratio} & \makecell[l]{Ratio of Lomb-Scargle score of \\ best two periods} \\
    \texttt{phase\_chisq} & \makecell[l]{$\chi^2$ of sinusoid fit to phase folded \\ lightcurve} \\
    \texttt{L\_m} & {Stetson L index} \\
    \texttt{J-H} & {2MASS J-H color} \\
    \texttt{J-K} & {2MASS J-K color} \\
    \texttt{W1-W2} & {WISE W1-W2 color} \\
    \texttt{W1-W4} & {WISE W1-W4 color} \\
    \texttt{W3-W4} & {WISE W3-W4 color} \\
\enddata
\end{deluxetable}

\subsection{Data Resampling}

To overcome the class imbalance in the initial training set, we use synthetic sampling methods implemented using the \texttt{imbalanced-learn} package \citep{Guillaume17}. Multiple strategies exist to selectively upsample the minority class and undersample the majority class. Based on trial and error, we choose the Adaptive Synthetic (ADASYN) sampling method \citep{Haibo2008} for upsampling and all K-Nearest Neighbours (allKNN) method \citep{Tomek76} for undersampling.

ADASYN sampling creates new synthetic samples for the minority class by interpolating between existing minority samples. This interpolation is performed by considering the neighboring minority samples and creating synthetic samples that lie along the line connecting them. ADASYN is unique in the sense that it adjusts the number and distribution of synthetic samples based on the difficulty of classification. It focuses on generating synthetic instances in the regions of the feature space that are challenging for the classifier. By doing so, it helps to balance the class distribution while maintaining the overall integrity of the data, resulting in improved model performance on imbalanced datasets.

The allKNN downsampling algorithm works by assigning a minority density to each sample in the minority class. The minority density is calculated as the number of minority samples in k-nearest neighbors, where ``k" is a user-controlled parameter. Samples from the majority class are selectively removed if they fall into a minority neighborhood with a density below a set threshold. Samples with density below the threshold could be sparsely located or overlap with the majority class, making it hard for the model to learn. This process is iterated till a desirable balance is achieved. allKNN downsampling, therefore helps create a balanced dataset while also removing samples that are harder to learn.

This combination is used to obtain a more balanced training set used to train the classifier.

%% file: ml_framework.tex
Decision trees and random forest-based classifiers have been shown to be effective for the task of classifying variable star lightcurves (see for e.g. \citealp{Sanchez2021}, \citealp{Boone2019}, \citealp{Masci14}, \citealp{Debosscher07}). Decision trees \citep{Breiman1984} are intuitive models having a tree-like branched node structure. They recursively partition the input data at each node to create homogeneous subgroups based on criteria set from the features. Data can be split either to maximize information gain or minimize impurity in partitioned sets. Complex decision trees may be prone to overfitting. Parameters such as the maximum depth of the tree and the minimum number of samples required to create a split can be controlled to avoid overfitting data. Random forests \citep{Breiman2001} is an ensemble method that builds multiple decision trees and combines their predictions to assign a probability to the predicted output. Different trees in the forest are built from random subsets of the training set and features. This helps avoid overfitting and gives improved performance. The importance of features in separating various classes in the input data can also be quantified in random forests.

We employ a gradient-boosted decision tree classifier implemented using the \texttt{LightGBM} framework \citep{Ke2017}. Gradient boosting is a popular technique for creating ensemble decision trees. It sequentially builds weak-learner trees where each tree is trained to correct the residuals of the preceding tree, ``boosting" its performance. Gradient descent is used to minimize a loss function, which is how the model learns. \texttt{LightGBM} offers an efficient implementation of gradient-boosted decision trees and reduces memory consumption by binning features.\\

For the training process, we split the bonafide dataset in a 70:30 train-test split. It is important that the model is not tested on resampled data, hence resampling is done only on 70\% of the bonafide set. LPVs are labelled as 1 and non-LPVs as 0. Due to class imbalance, the overall accuracy of the classifier may be misleading. We use the following metrics to quantify model performance:

\vspace{-0.4cm}
\begin{center}
\begin{equation}
    \textrm{Recall / True positive rate (TPR)} = \frac{\textrm{TP}}{\textrm{TP} + \textrm{FN}}
\end{equation}
\end{center}
\vspace{-0.75cm}
\begin{center}
\begin{equation}
    \textrm{True negative rate (TNR)} = \frac{\textrm{TN}}{\textrm{TN} + \textrm{FP}}
\end{equation}
\end{center}
\vspace{-0.75cm}
\begin{center}
\begin{equation}
    \textrm{Precision (prec)} = \frac{\textrm{TP}}{\textrm{TP} + \textrm{FP}}
\end{equation}
\end{center}
\vspace{-0.75cm}
\begin{center}
\begin{equation}
    \textrm{F-measure} = \frac{2*\textrm{prec}*\textrm{recall}}{\textrm{prec} + \textrm{recall}}
\end{equation}
\end{center}
\vspace{-0.75cm}
\begin{center}
\begin{equation}
    \textrm{Weighted g-mean} = (\textrm{TPR}^{\textrm{P}} * \textrm{TNR}^{\textrm{N}})^{\frac{1}{\textrm{P}+\textrm{N}}}
\end{equation}
\end{center}

where TP is true positive, FP is false positive, TN is true negative, FN is false negative, P is all positives, and N is all negatives.\\
The true positive rate is the probability of an LPV being labeled correctly, and the true negative rate is the probability of a non-LPV being labeled correctly. Precision gives the fraction of all LPVs classified correctly. The F-measure is an accuracy metric given by the harmonic mean of precision and recall. The g-mean is the geometric mean of TPR and TNR. We observe that due to class imbalance in the test set, the simple geometric-mean tends to favor models that have a lower misclassification rate of non-LPVs. To mitigate this bias, we calculate the weighted geometric mean of TPR and TNR, using the total samples in each class as weights to get a robust measure of classifier performance.

\subsection{Training on the bonafide set}

\begin{figure}
\epsscale{1.2}
\plotone{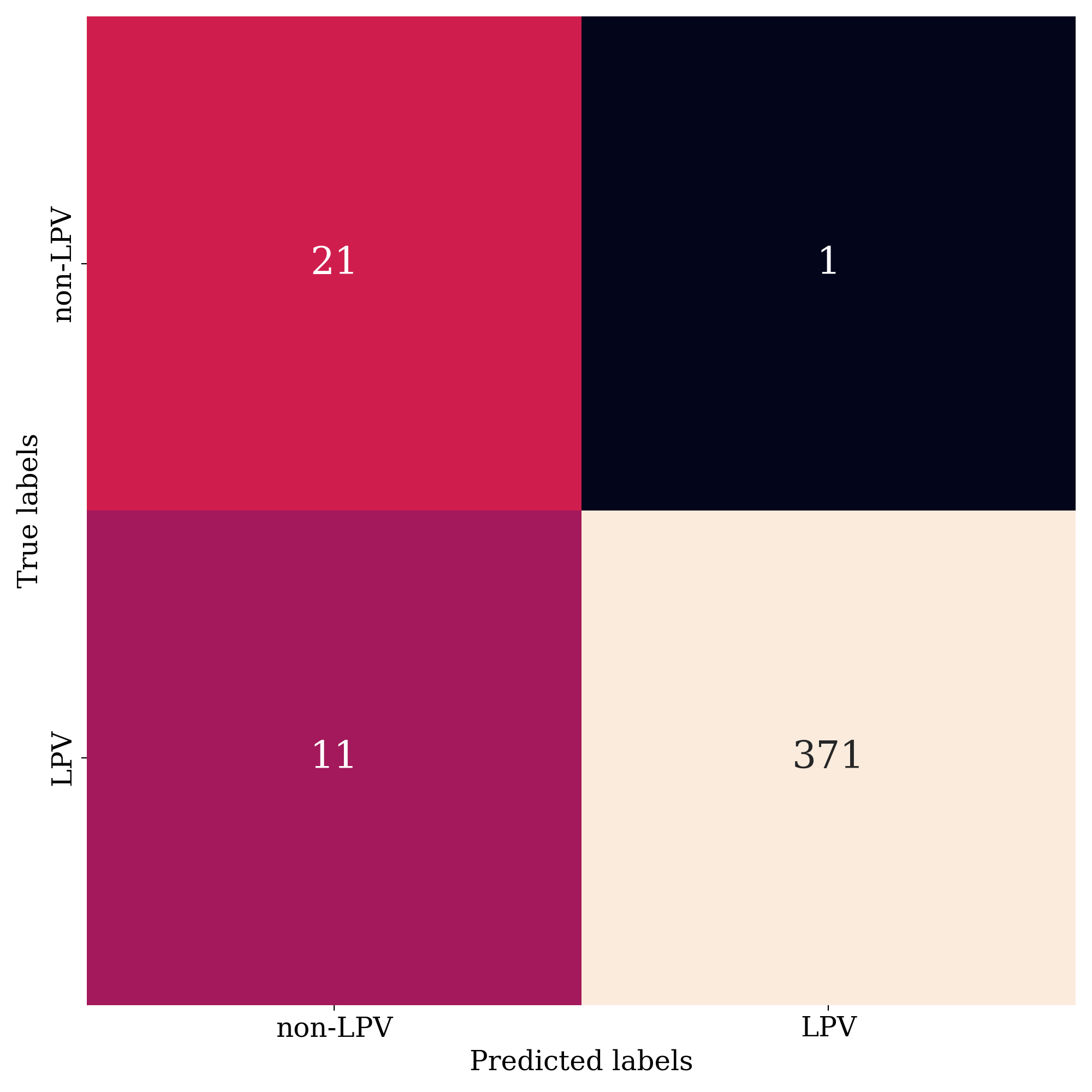}
\caption{Confusion matrix from training on the bonafide dataset. This model will be referred to as LGBMv0.1. Metrics: TPR = 97.120\%, TNR =
95.455\%, prec = 99.731\%, F-measure = 0.984, weighted g-mean = 0.970
\label{fig:cm_first}}
\end{figure}

The \texttt{LightGBM} gradient-boosted decision tree has a large number of hyperparameters that need to be tuned to get the most out of the classifier. A grid search that chooses the optimal hyperparameters by trying each possible combination would be time-consuming and is not feasible. In each training iteration, we randomly sample 200 points from the hyperparameter space and choose the set with the highest accuracy. This method has been shown to be effective at achieving near-optimal performance with high probability \citep{Bergstra12}. For each combination of hyperparameters, accuracy is calculated using five-fold cross-validation. The training dataset is partitioned into five folds, with the model being fit to four such folds and evaluated on the remaining fold. This process is repeated to effectively sample the training data. We evaluate the metrics on the model and create a confusion matrix. For binary classification tasks, the confusion matrix is a 2 x 2 matrix showing the distribution of true negatives, false positives, false negatives, and true positives. Examining the confusion matrix gives good insights into the model's performance, as all metrics are derived from its quantities. Figure \ref{fig:cm_first} shows the confusion matrix from training on the bonafide dataset.

\begin{figure*}[!ht]
\epsscale{1.2}
\plotone{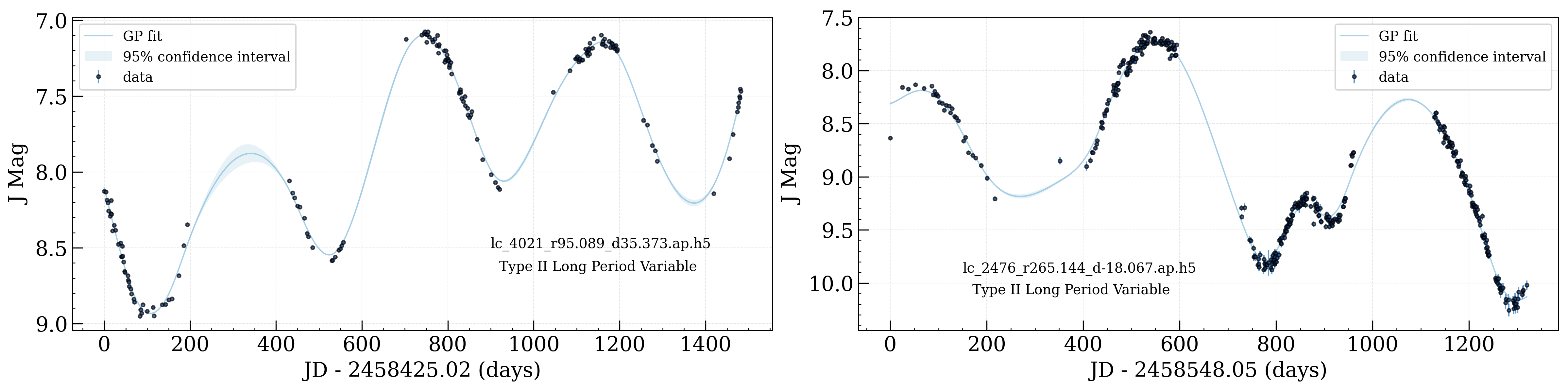}
\caption{Examples of LPVs classified as Type II LPVs. Type II LPVs display sinusoidal evolution, however, additional trends such as a constant rise or deacy, varying amplitudes of oscillation or presence of several oscillation frequencies are also present. 
\label{fig:example_typeii_lpv}}
\end{figure*}

Following this training procedure, it is important to understand samples misclassified by the model. We save the predicted probabilities and labels for all samples in the training set and compare labels with the true values. Lightcurves of all misclassified samples are plotted and saved along with the confusion matrix. Since we perform a randomized search for hyperparameters, we run multiple training iterations and pick the best-performing model. Examination of confusion matrix and misclassified samples is particularly helpful in choosing the optimal model in cases where certain samples may be mislabelled in the training set. This is particularly unavoidable when we build the training set in an iterative process and predictions from initial models form the true labels of the next iteration.

\subsection{Expanding the training set}
\label{sec:expanded_training}
The first step in expanding the training dataset is to choose a larger set of data to obtain predictions on. We extracted features for 35 million lightcurves in the PGIR lightcurve database. Of these, we choose a subsample of $\sim$12000 sources by applying cuts on the von Neumann score ($\eta$ < 1.5), peak-to-peak amplitude (ptp > 1) and Lomb-Scargle score (LSscore > 0.75). We run the LGBMv0.1 classifier on all of these sources and choose only those samples with a prediction probability >0.9, resulting in a larger training set of 9815 LPVs and 443 non-LPVs. 

However, upon visual inspection of the classifications, we find that the misclassification rate is high in this dataset. We observe that a large fraction of the classified non-LPVs are LPVs that show variations in amplitude, show multiple pulsation modes or have a slow background decay or rise in addition to the few hundred day periodic variations. Upon training on the extended dataset with corrected labels, we observe that the classifier struggles to separate these erratic LPVs, generally deviating from a clean sinusoidal evolution, from non-LPVs. Hence we create a third category -- ``Type II LPVs", based purely on lightcurve morphology comprising of such non-sinusoidal LPVs. We note that this classification does not necessarily hold a physical meaning, as many AGB stars typically do display such evolution and can have multiple simultaneous periodic excitations. However, separating the LPVs into cleanly sinusoidal and erratic evolution better enables the classifier to pick up the general features of each class. We combine the sinusoidal LPVs and type II LPVs into the LPV class following classification. Representative examples of type II LPVs are shown in Figure \ref{fig:example_typeii_lpv}.

\begin{figure}
\epsscale{1.2}
\plotone{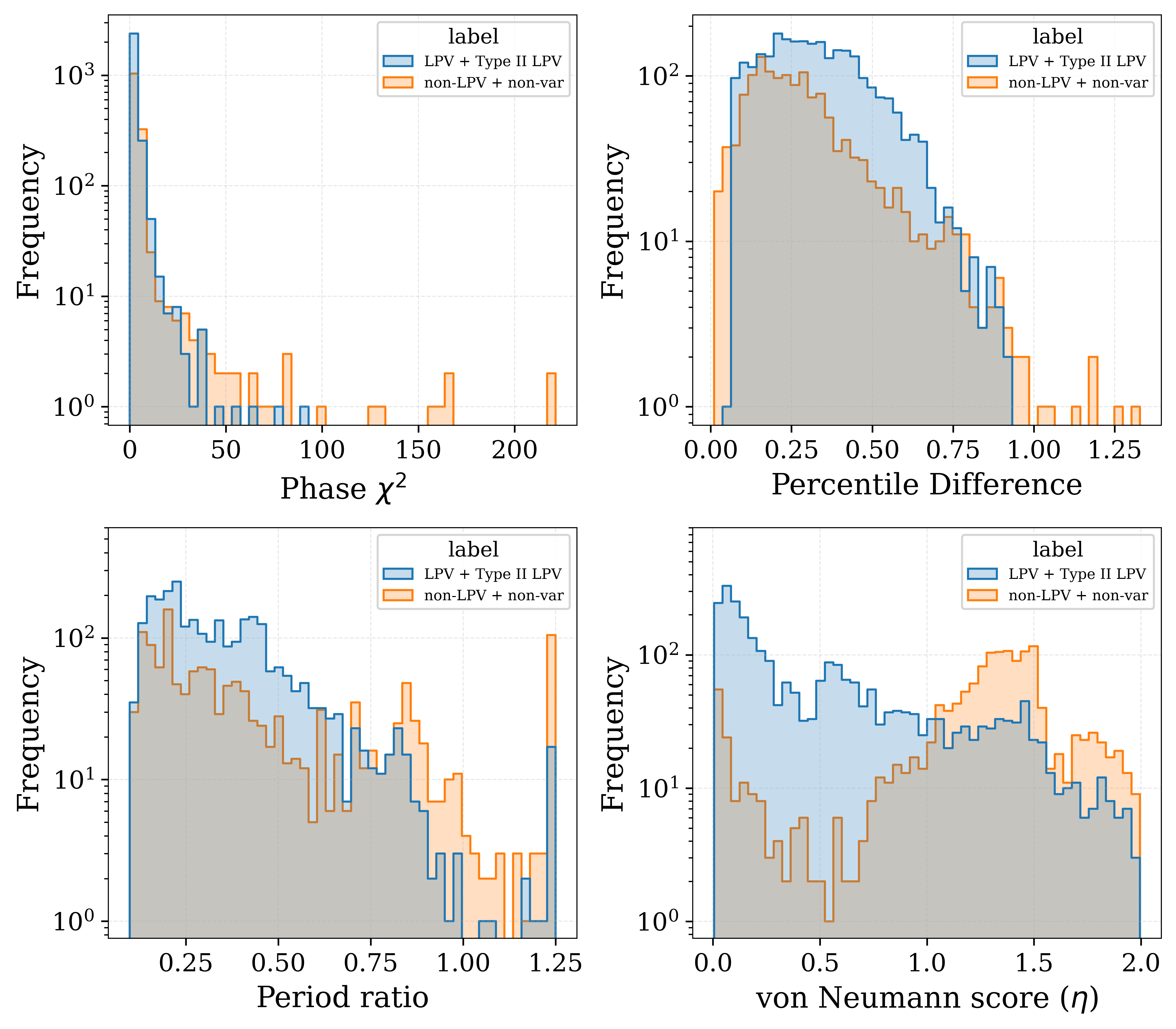}
\caption{Distributions of select features for variables in the expanded training set. We observe that the features occupy similar values for the classes shown here as the classifier has to separate low amplitude variables whose feature phase space is not too different from non-variables.
\label{fig:features_hist_final}}
\end{figure}

\begin{figure}
\epsscale{1.0}
\plotone{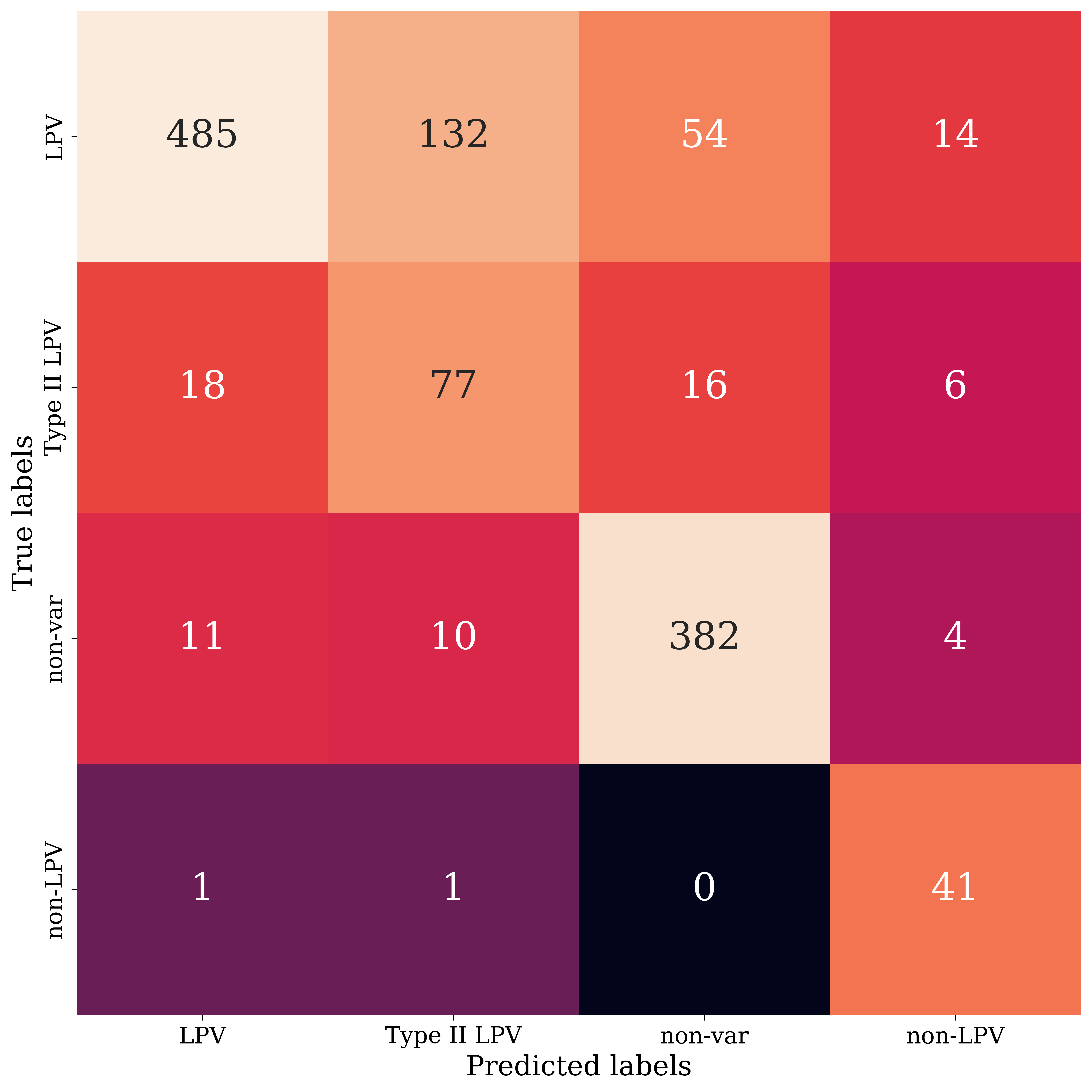}
\caption{Confusion matrix from training on the extended labeled set. This model will be referred to as LGBMv0.2. Since we classify between three classes, we calculate metrics for each class. Metrics (listed for LPVs, Type II LPVs, non-vars and non-LPVs respectively): TPR = 70.80\%,  65.81\%, 93.86\%, 95.35\%, TNR = 94.71\%, 87.40\%, 91.72\%, 98.02\%, precision = 94.12\% 35.0\%, 84.51\%, 63.08\%, F-measure = 0.81, 0.46, 0.89, 0.76 and weighted g-mean = 0.80, 0.69, 0.93, 0.95
\label{fig:cm_second}}
\end{figure}

We repeat the training procedure on this larger sample of labeled data. In this iteration, we observe that the classifier has improved performance and is better able to distinguish the type II LPVs and non-LPVs, as shown in the confusion matrix in Figure \ref{fig:cm_second}. We assemble a clean set of non-vars by visually inspecting a subset of lightcurves chosen to have von Neumann score less than 1.5, amplitude less than 1 mag and LPV probability less than 0.3 from the classifier trained on bonafide variables. Finally, we build a master training set consisting of 2335 LPVs, 444 Type-II LPVs, 1332 non-vars and 166 non-LPVs. The distribution of select features for LPVs and non-LPVs in the bonafide set is shown in Figure \ref{fig:features_hist_final}. Several training iterations of the model on this training set resulted in a best-performing model with a g-mean score of 0.95. 

\begin{figure}
\epsscale{1.0}
\plotone{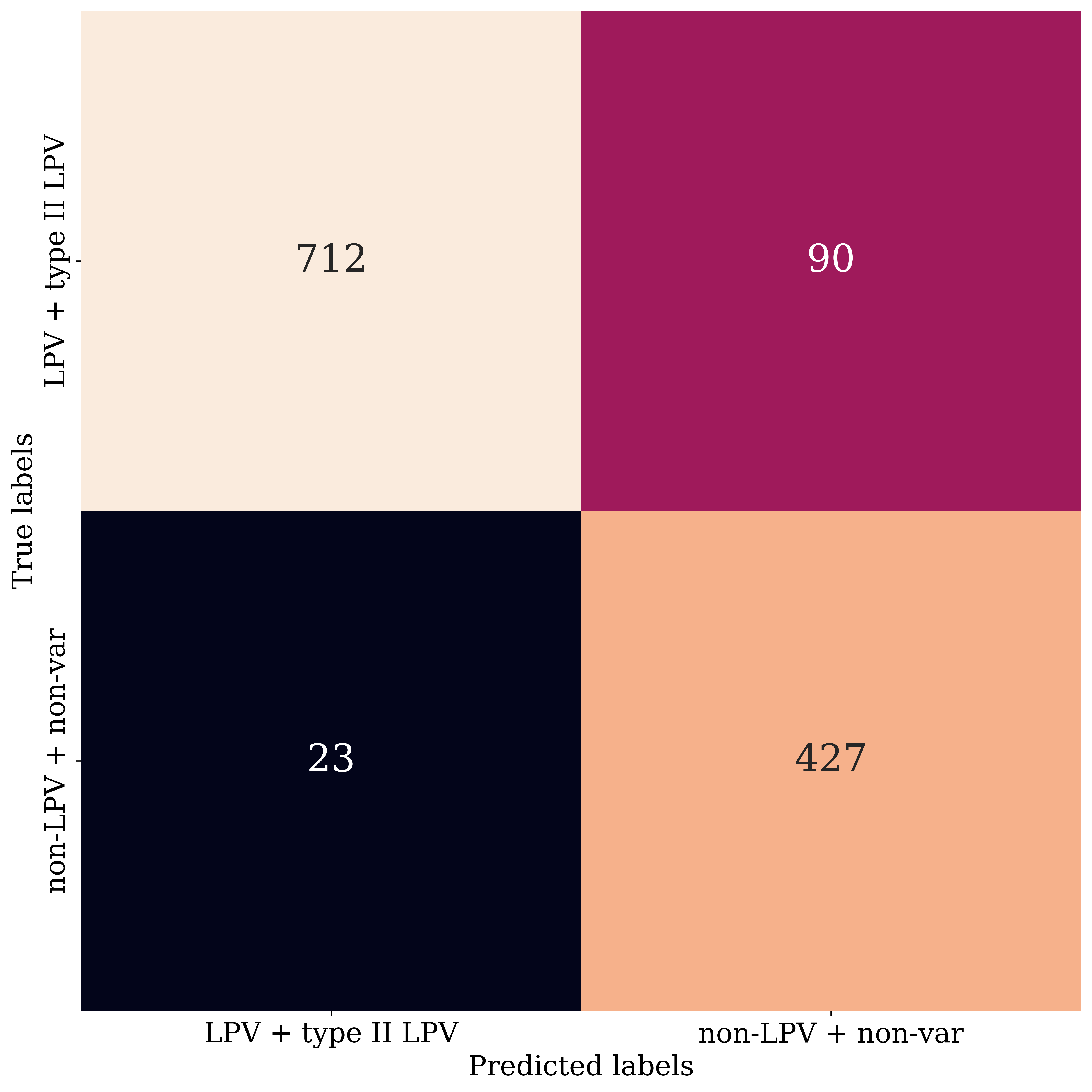}
\caption{Confusion matrix from training on the extended labeled set with Type II LPVs grouped with LPVs and non-LPVs grouped with non-vars. Metrics: TPR = 94.89\%, TNR = 88.78\%, precision = 82.56\%, F-measure = 0.88, weighted g-mean = 0.95
\label{fig:cm_LPV_second}}
\end{figure}

The model shows high confusion between LPVs and Type II LPVs, but is able to minimize confusion between the combined set of these two and the combined set of non-vars and non-LPVs. We sum the probability score of LPV and type II LPV is to calculate the net probability of a source being an LPV. This allows us to separate LPVs from non-LPVs and non-vars in an efficient manner. We group all of the type II LPVs into the LPV class and the non-vars into the non-LPV class to obtain the final confusion matrix and metrics shown in Figure \ref{fig:cm_LPV_second}. We also list the optimal parameters of the gradient-boosted classifier in Table \ref{tab:hyperparameters}.

\subsection{Feature Importance}

\begin{figure*}
\epsscale{1.0}
\plotone{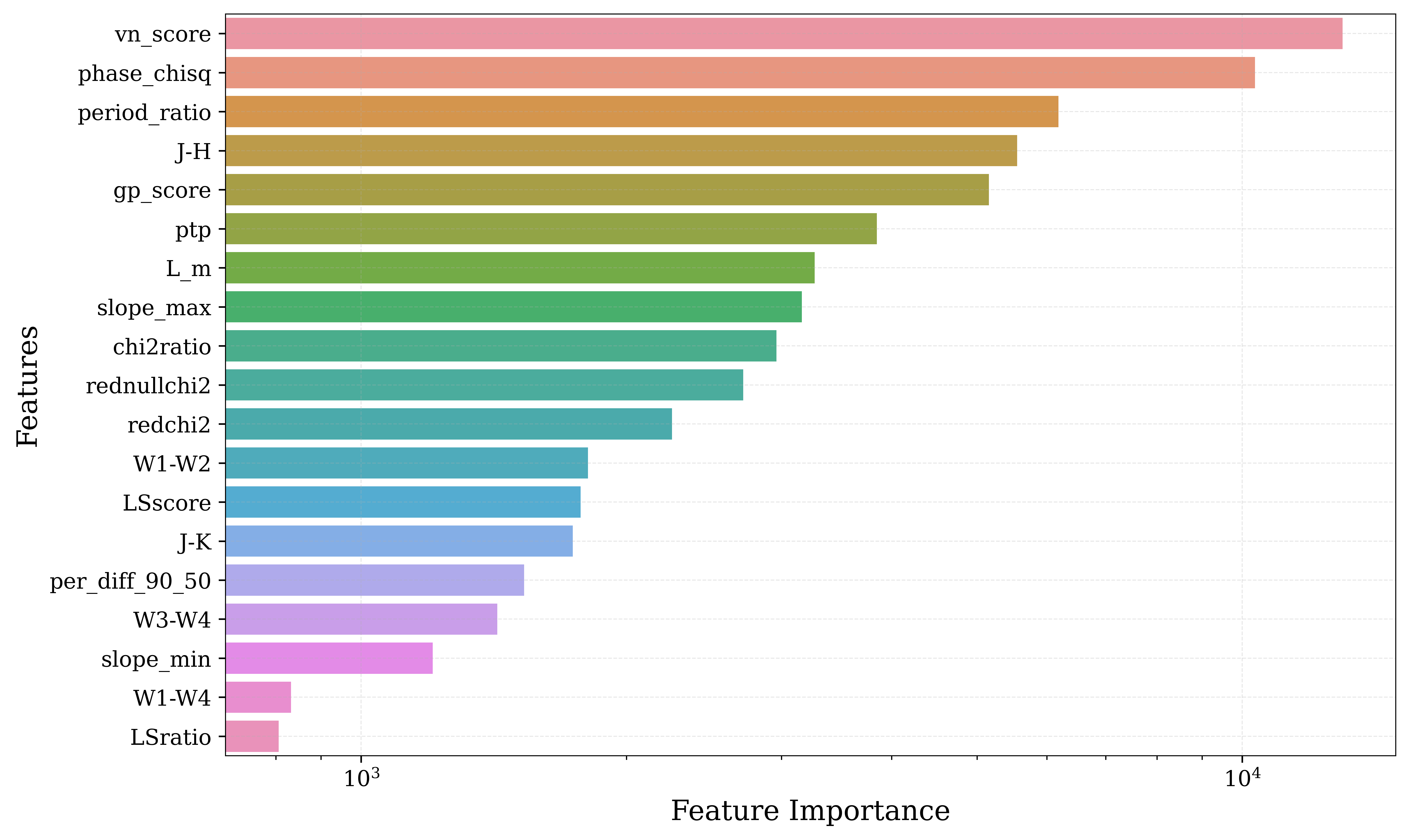}
\caption{Feature importance of various features used in training. The feature importance quantifies how effective a particular feature is in separating an LPV and a non-LPV. We notice features such as the von Neumann score and period ratio are very effective, while the Lomb-Scargle score is not as useful for the classifier.
\label{fig:feat_imp}}
\end{figure*}

Tree-based classifiers provide convenient methods to quantify the relative importance of different features used in training. The \texttt{LightGBM} framework, in particular, calculates feature importance in two ways, using ``split" and ``gain". Split-wise importance is calculated as the number of times a particular feature is used in the model to partition data. Since the model tries to split data to minimize a loss function, the number of times a feature is used in this process is a good measure of its effectiveness. Gain-wise importance calculates the impurity reduction in each split; impurity reduction is the difference in loss of parent tree node and the weighted sum of loss of child tree nodes after the split. This quantifies how much a split improves the model's performance. Impurity reduction for each feature is accumulated for all trees in the ensemble and normalized to obtain the final relative importance. We adopt gain-based importance as the preferred metric as it provides a direct measure of improvement in accuracy contributed by a feature.

Figure \ref{fig:feat_imp} shows the calculated feature importance. We note that the period ratio is among the most important features for the classifier through both the gain and split importance calculations. This is consistent with other variable star or transient alert classifiers (e.g.  \citealp{Sanchez2021}, \citealp{Masci17}, \citealp{Kim16}). The $\chi^2$ of a sinusoid fit to the phase folded lightcurve and the J-H color also rank highly in both metrics, while the Stetson L index and peak-to-peak amplitude perform decently. We also observe that the Lomb-Scargle scores and percentile difference perform poorly in separating LPVs and non-LPVs.

\begin{deluxetable}{ll}
\epsscale{0.8}
\tablecaption{Optimal hyperparameters of the \texttt{LightGBM} classifier.
\label{tab:hyperparameters}}
\tablehead{
    \colhead{Hyperparameter} &
    \colhead{Value}
}
\startdata
    \texttt{objective} & \texttt{multiclass} \\
    \texttt{boosting\_type} & \texttt{gbdt} \\
    \texttt{colsample\_bytree} & 0.9\\
    \texttt{max\_depth} & 9\\
    \texttt{min\_child\_samples} & 40\\
    \texttt{min\_split\_gain} & 1\\
    \texttt{n\_estimators} & 460\\
    \texttt{num\_leaves} & 197\\
    \texttt{reg\_alpha} & 0.035\\
    \texttt{reg\_lambda} & 0.03\\
    \texttt{subsample} & 0.90\\
    \texttt{subsample\_freq} & 59\\
\enddata
\end{deluxetable}

%% file: catalog.tex
\begin{figure*}[!ht]
\epsscale{1.2}
\plotone{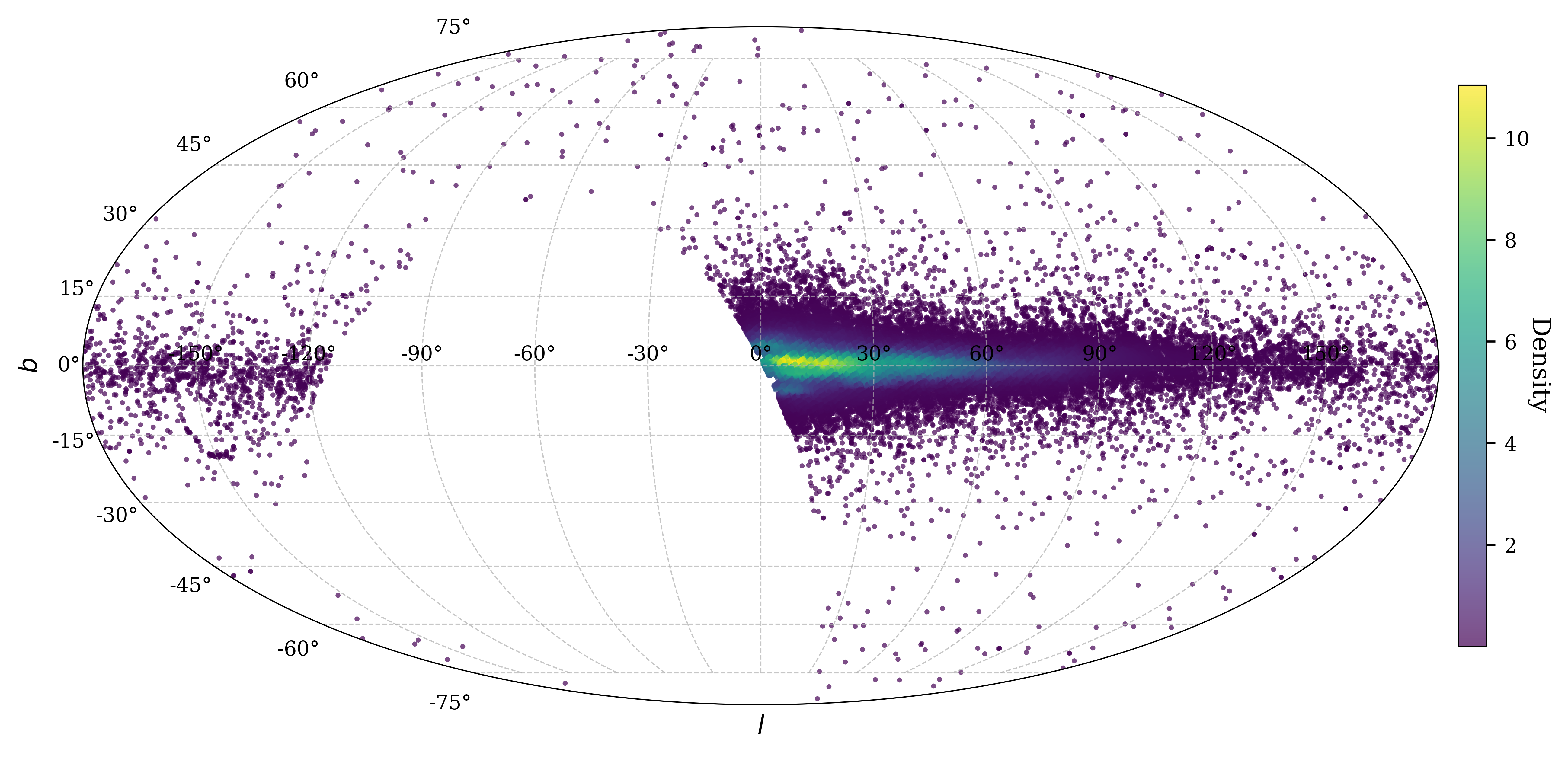}
\caption{Spatial distribution of long-period variables from Gattini. A large fraction (152,220) of LPVs from the catalog lie within the galactic latitudes of $-10^{\circ}$ to $+10^{\circ}$. The missing patch of sky is below a declination of $-28^{\circ}$, which PGIR cannot observe. The color bar represents the density of stars, with a high density around the galactic centre.
\label{fig:lpv_spatial_dist}}
\end{figure*}

\begin{figure*}
\epsscale{1.2}
\plotone{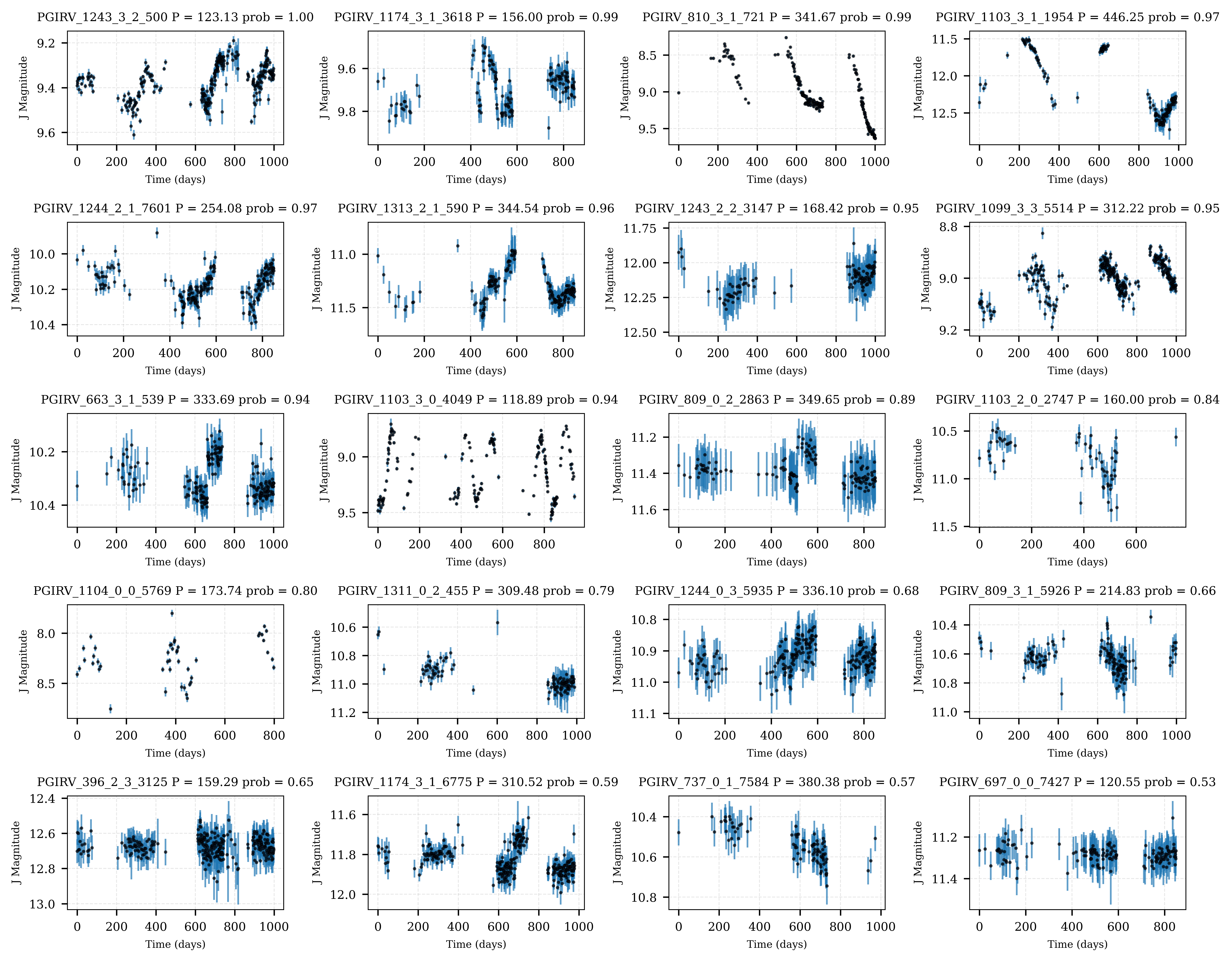}
\caption{Example lightcurves of 20 LPVs, with the periods and LPV probabilites from our classifier. We see a good mix of large amplitude and small amplitude LPVs, showing a variety of morphologies. The period and internal designation of each LPV is listed. Lightcurves for all LPVs from the catalog are publicly available. A few suspicious lightcurves can also be seen in the above sample, which have a low classifier confidence (typically less than 0.6).
\label{fig:lpv_lcs}}
\end{figure*}

We use the LGBMv0.2 classifier to predict the classes of 35 million PGIR lightcurves, chosen to have a von Neumann score less than 2, and requiring at least 30 detections. We present a catalog of 159,696 stars having $\textit{prob\_lpv} > 0.5$ from the PGIR survey in Table \ref{tab:lpvcat}, containing the following columns. ``Name" refers to the internal designation of the variable, ``RA" and ``Dec" give the spatial position, ``num\_detections" is the number of detections, ``lcdur" is the baseline of observation, given by the difference of the time of final detection and the time of initial detection, ``bestperiod" and ``amplitude" are the period and amplitude, ``$\left<J\right>$" is the mean J band magnitude, ``prob\_lpv" is the classifier confidence. In addition, all extracted features and 2MASS and WISE color information are also presented. 
Figure \ref{fig:lpv_lcs} shows example lightcurves of 20 LPVs from the catalog.

\begingroup
\renewcommand{\tabcolsep}{2pt}
\begin{table*}[ht]
\begin{center}
\begin{minipage}{20cm}
    \begin{threeparttable}
    \begin{tabular}{lcccccccc}
    \toprule
    Name & RA & Dec & num\_dets & lcdur & amplitude & bestperiod & $\left<J\right>$ & prob\_lpv \\
    \midrule
    PGIRV\_1008\_0\_3\_357  & 172.238841 & -7.260807  & 83  & 962.6 & 0.07 & 130 & 9.41  & 0.86 \\
    PGIRV\_1012\_0\_1\_365  & 190.461586 & -7.169434  & 64  & 915.6 & 0.15 & 378 & 14.91 & 0.73 \\
    PGIRV\_1012\_2\_0\_355  & 192.904506 & -8.410068  & 68  & 915.6 & 0.18 & 409 & 13.03 & 0.50 \\
    PGIRV\_1015\_0\_2\_1656 & 205.447468 & -8.229639  & 69  & 892.6 & 0.12 & 155 & 12.87 & 0.96 \\
    PGIRV\_1017\_3\_2\_1044 & 218.119784 & -5.521623  & 51  & 885.6 & 1.26 & 276 & 13.98 & 0.93 \\
    PGIRV\_1018\_1\_3\_1424 & 220.276317 & -4.953930  & 66  & 904.6 & 0.05 & 240 & 11.37 & 0.52 \\
    PGIRV\_1049\_2\_1\_115  & 13.338281  & -11.976007 & 90  & 989.1 & 0.20 & 195 & 11.47 & 0.99 \\
    PGIRV\_1054\_0\_1\_194  & 35.383878  & -12.128101 & 96  & 838.8 & 0.07 & 173 & 10.12 & 0.70 \\
    PGIRV\_1084\_1\_3\_1287 & 183.570633 & -9.013661  & 83  & 934.6 & 0.60 & 159 & 13.22 & 0.92 \\
    PGIRV\_1086\_0\_3\_472  & 194.237602 & -11.771883 & 84  & 943.6 & 0.20 & 236 & 10.73 & 0.94 \\
    PGIRV\_1091\_3\_1\_1781 & 219.437017 & -9.584581  & 73  & 909.6 & 0.07 & 229 & 10.12 & 0.62 \\
    PGIRV\_1119\_1\_0\_0    & 354.895573 & -10.027897 & 71  & 975.2 & 0.07 & 188 & 10.13 & 0.52 \\
    PGIRV\_1119\_3\_1\_1517 & 357.465153 & -9.885638  & 90  & 991.1 & 0.15 & 353 & 11.44 & 0.51 \\
    PGIRV\_1120\_2\_2\_851  & 4.066668   & -17.603277 & 85  & 981.2 & 0.59 & 132 & 11.52 & 0.88 \\
    PGIRV\_1121\_3\_0\_1348 & 7.516286   & -15.770552 & 17  & 642.2 & 1.06 & 102 & 15.75 & 0.54 \\
    PGIRV\_1122\_3\_1\_364  & 13.373073  & -14.509416 & 88  & 989.2 & 0.27 & 175 & 10.64 & 0.99 \\
    PGIRV\_1123\_0\_0\_73   & 15.983818  & -17.514027 & 97  & 993.1 & 0.14 & 158 & 11.05 & 0.93 \\
    PGIRV\_1125\_0\_3\_507  & 26.950635  & -16.721510 & 42  & 784.8 & 0.22 & 241 & 16.53 & 0.55 \\
    PGIRV\_1130\_2\_0\_1225 & 52.649587  & -17.941429 & 88  & 846.7 & 0.16 & 103 & 15.87 & 0.68 \\
    PGIRV\_1157\_1\_1\_652  & 185.468635 & -13.886072 & 81  & 893.6 & 0.69 & 222 & 8.659 & 0.92 \\
    PGIRV\_1190\_1\_0\_597  & 350.546129 & -15.304388 & 86  & 989.2 & 0.37 & 159 & 10.86 & 0.90 \\
    PGIRV\_1191\_1\_1\_948  & 355.128686 & -13.872989 & 85  & 970.1 & 0.05 & 182 & 9.748 & 0.57 \\
    PGIRV\_1192\_2\_0\_573  & 3.142701   & -22.921270 & 89  & 988.2 & 0.78 & 170 & 10.39 & 0.97 \\
    PGIRV\_1198\_1\_3\_1166 & 32.186023  & -19.215745 & 10  & 751.9 & 0.80 & 256 & 15.05 & 0.55 \\
    PGIRV\_1199\_1\_3\_621  & 37.821288  & -19.131971 & 104 & 838.7 & 0.19 & 236 & 10.40 & 0.98 \\
    PGIRV\_1199\_2\_3\_1245 & 39.979909  & -21.072354 & 10  & 364.0 & 0.42 & 125 & 15.76 & 0.57 \\
    PGIRV\_1256\_2\_3\_445  & 333.906426 & -22.015062 & 13  & 994.2 & 1.09 & 361 & 16.20 & 0.84 \\
    PGIRV\_1258\_1\_2\_436  & 341.508786 & -20.476088 & 76  & 991.2 & 0.37 & 135 & 12.57 & 0.68 \\
    PGIRV\_1260\_1\_1\_910  & 349.897992 & -18.939772 & 78  & 974.2 & 0.81 & 151 & 11.50 & 0.94 \\
    PGIRV\_1260\_2\_1\_184  & 353.313861 & -21.525274 & 82  & 974.2 & 0.23 & 114 & 11.18 & 0.75 \\
    PGIRV\_1263\_3\_3\_824  & 9.703086   & -23.926492 & 76  & 786.8 & 0.59 & 177 & 11.21 & 0.99 \\
    PGIRV\_1266\_2\_3\_1063 & 25.300664  & -26.278947 & 41  & 788.8 & 0.24 & 232 & 16.48 & 0.58 \\
    PGIRV\_126\_1\_1\_1490  & 179.680170 & 63.354066  & 144 & 977.6 & 0.25 & 139 & 10.43 & 0.64 \\
    PGIRV\_1296\_1\_0\_1458 & 180.104200 & -24.686672 & 29  & 837.7 & 0.59 & 175 & 10.18 & 0.58 \\
    PGIRV\_1296\_1\_1\_58   & 181.156892 & -24.160596 & 80  & 876.6 & 0.27 & 139 & 11.24 & 0.90 \\
    PGIRV\_1299\_1\_1\_373  & 196.753528 & -24.091946 & 73  & 877.6 & 0.45 & 162 & 9.50  & 0.99 \\
    PGIRV\_129\_1\_1\_808   & 209.470908 & 64.111660  & 141 & 930.7 & 0.69 & 138 & 11.77 & 0.85 \\
    PGIRV\_130\_2\_2\_1558  & 225.541293 & 59.522531  & 145 & 932.7 & 0.32 & 373 & 8.530 & 0.71 \\
    PGIRV\_1313\_0\_3\_2444 & 271.707078 & -26.502290 & 34  & 735.9 & 0.25 & 149 & 11.52 & 0.87 \\
    PGIRV\_1324\_1\_3\_418  & 330.479227 & -23.974536 & 66  & 994.2 & 0.06 & 162 & 9.349 & 0.73 \\
    PGIRV\_1325\_0\_1\_761  & 334.291335 & -26.117588 & 68  & 987.2 & 0.88 & 341 & 11.04 & 0.93 \\
    PGIRV\_1325\_3\_3\_1070 & 337.938061 & -23.546005 & 69  & 987.2 & 0.75 & 209 & 11.48 & 0.92 \\
    PGIRV\_1326\_0\_1\_412  & 339.714107 & -26.934976 & 60  & 988.2 & 0.31 & 130 & 11.57 & 0.92 \\
    \bottomrule
    \end{tabular}
    \begin{tablenotes}
        \small{
        \item Description of columns :\\
        $Name :$ PGIR name for variable; 
        $RA$ and $Dec :$ RA and Dec (in degrees for equinox 2000);
        $num\_detections:$ Number of detections; \\
        $lcdur :$ Observation baseline;
        $amplitude$ and $bestperiod :$ amplitude and period;
        $\left<J\right>$: mean J band magnitude;
        $prob\_lpv :$ \\ Classifier confidence; \\
        The complete catalog will be available as a machine-readable table.}
    \end{tablenotes}
    \end{threeparttable}
    \caption{PGIR Long Period Variables}
    \label{tab:lpvcat}
    \end{minipage}
    \end{center}
\end{table*}
\endgroup

\subsection{Catalog Features}

\begin{figure}
\epsscale{1.2}
\plotone{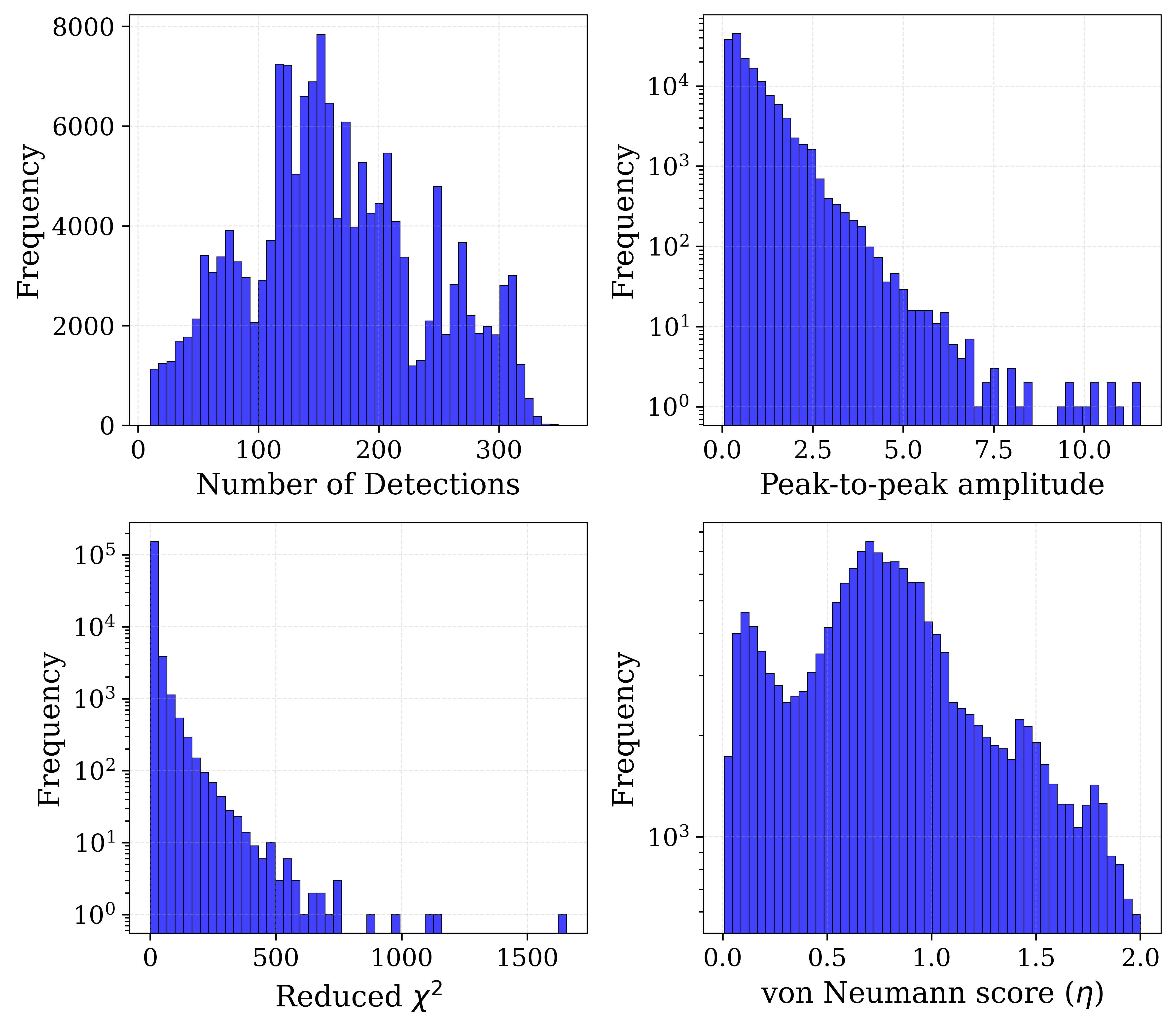}
\caption{Distribution of select features in the LPV catalog. The majority of LPVs in the catalog (118,112) have amplitude less than 0.25 mag and are likely to be semiregular variables. The von Neumann score shows a bimodal distribution with peaks corresponding to Miras and semiregular variables.
\label{fig:cat_features}}
\end{figure}

Figure \ref{fig:lpv_spatial_dist} shows the spatial distribution of LPVs. We observe that 155,220 of LPVs are located between galactic latitudes ($b_{gal}$) of $-10^{\circ}$ to $+10^{\circ}$. We observe that a large number of LPVs have low amplitudes $(<0.25)$ mag and a low percentile difference, implying that they are likely to be semi-regular variables, while the LPVs showing amplitudes greater than 0.5 mag are likely to be Miras. Figure \ref{fig:cat_features} shows the distribution of the peak-to-peak amplitude, the number of detections for each variable, the von Neumann score ($\eta$), and the reduced $\chi^2$ of a sinusoid fit to a lightcurve. 

\begin{figure}
\epsscale{1.2}
\plotone{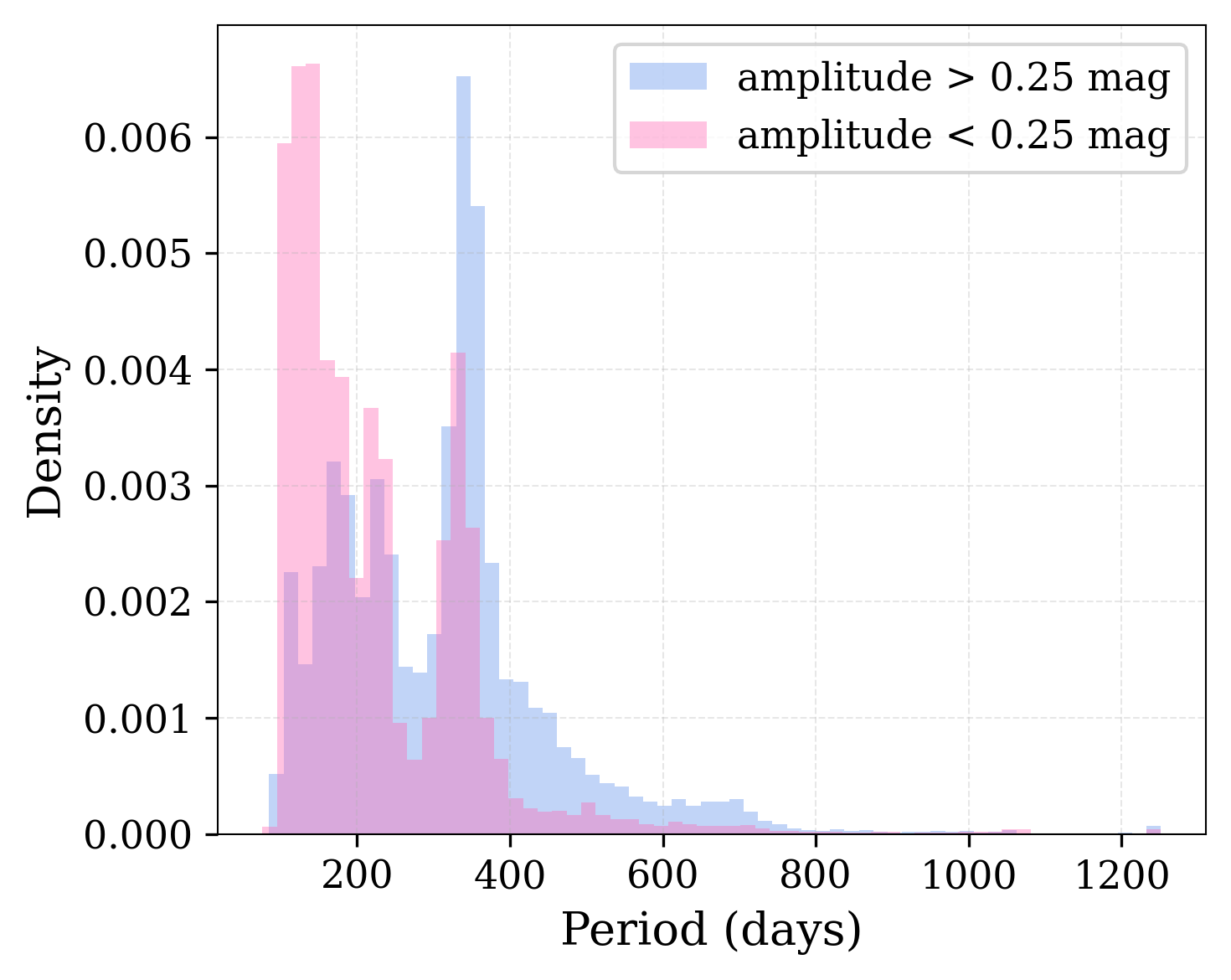}
\caption{Distribution of periods in the PGIR LPV catalog, separated by amplitude, representing Miras (blue histogram) and semiregular variables (pink histogram). We show the density in each bin. We observe a bimodal period distribution for Miras with peaks at \til200 d and \til340 d, while the semiregular variables show peaks at \til130 d, \til220 d, and \til330 d.
\label{fig:period_dist}}
\end{figure}

The period histogram of LPVs is shown in Figure \ref{fig:period_dist}, with the Miras showing a doubly peaked distribution and the semiregular variables showing a triply peaked distribution. We obtain peaks at \til200 d and \til340 d for the Miras, consistent with other studies \citep{Matsunaga05, Matsunaga09, Glass01, Nikzat22}, which also find bimodal period distributions for Miras. The semiregular variables show peaks in their period distribution at \til130 d, \til220 d, and \til330 d. \\
\citet{Conroy15} introduced a period-amplitude relationship for Miras:

\begin{equation}\label{eq:2}
\log_{10}{\Delta{J}} = \alpha \log_{10}P + \beta
\end{equation}

\begin{figure}
\epsscale{1.2}
\plotone{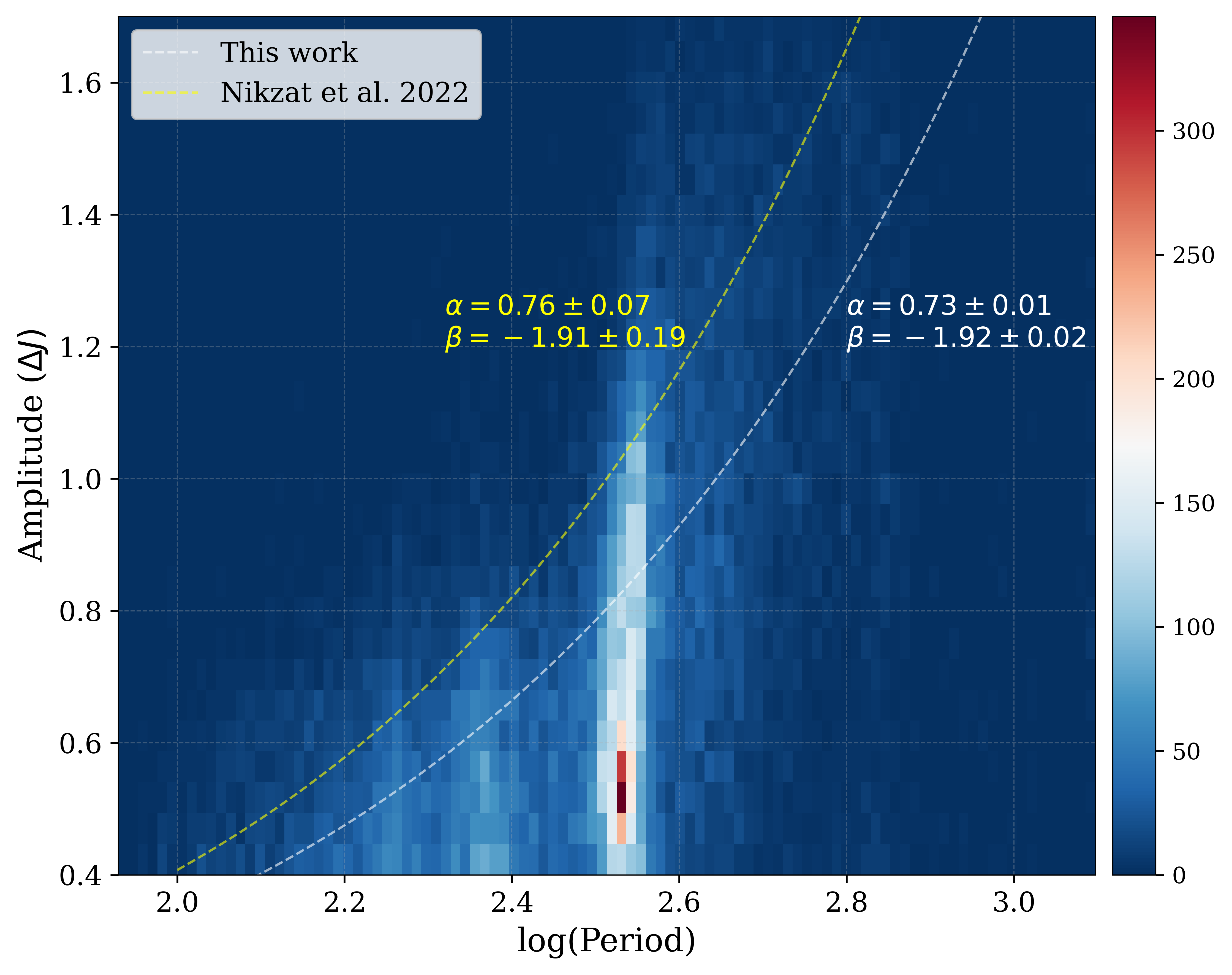}
\caption{Period Amplitude relationship of Mira variables, chosen to be LPVs from the catalog having amplitude greater than 0.4 mag. The white curve represents the best-fit parameters from PGIR Miras, compared to the relation obtained by \citet{Nikzat22}. While the general trend of the period-amplitude relation holds, there is a significant scatter around the best-fit curve. The distribution of PGIR Miras in the log(period)-amplitude space is shown as a 2D histogram with the colorbar representing the number of objects in each bin.
\label{fig:pl}}
\end{figure}

We examine this relation for the Mira variables from PGIR in the J-band as shown in Figure \ref{fig:pl} and compare it to the relation obtained in \citet{Nikzat22}. While the general trend holds, there is a significant scatter around the best-fit curve. We obtain $\alpha = +0.727 \pm 0.007$ and $\beta = -1.923 \pm 0.019$ from Miras in PGIR data. Our best-fit parameters agree reasonably with \citet{Nikzat22}, with $\alpha = +0.76 \pm 0.07$ and $\beta = -1.91 \pm 0.19$.


\subsection{Catalog Validation}

\begin{figure}
\epsscale{1.2}
\plotone{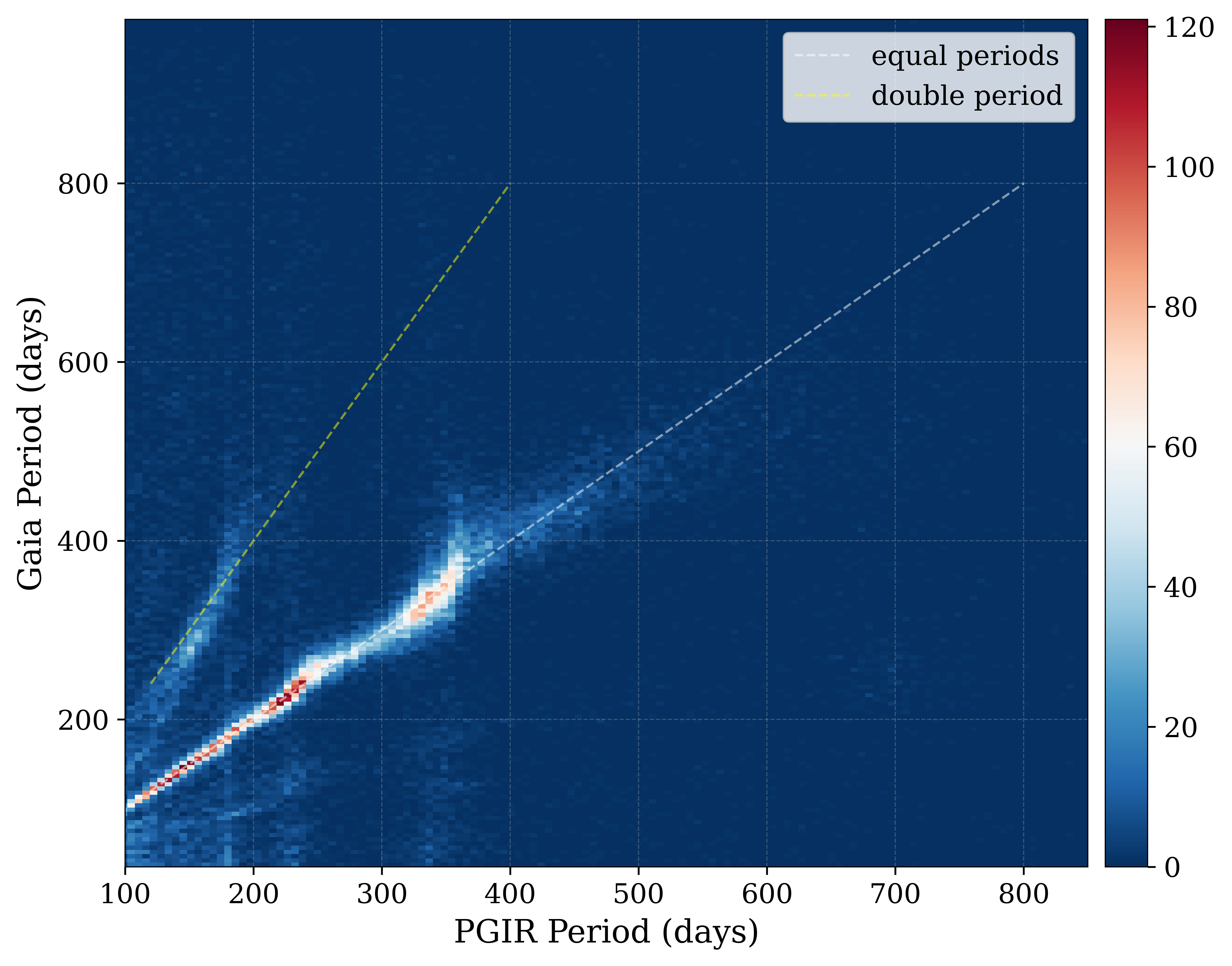}
\caption{Comparison between periods of crossmatched LPVs from PGIR and Gaia. We observe a strong correlation between the two datasets. We also notice a secondary branch, which approximately represents the line where the period from Gaia is approximately twice the period from PGIR. The color gradient represents the number of objects in each bin of the 2D histogram.
\label{fig:gaia_period_xmatch}}
\end{figure}

\begin{figure*}[!ht]
\epsscale{1.0}
\plotone{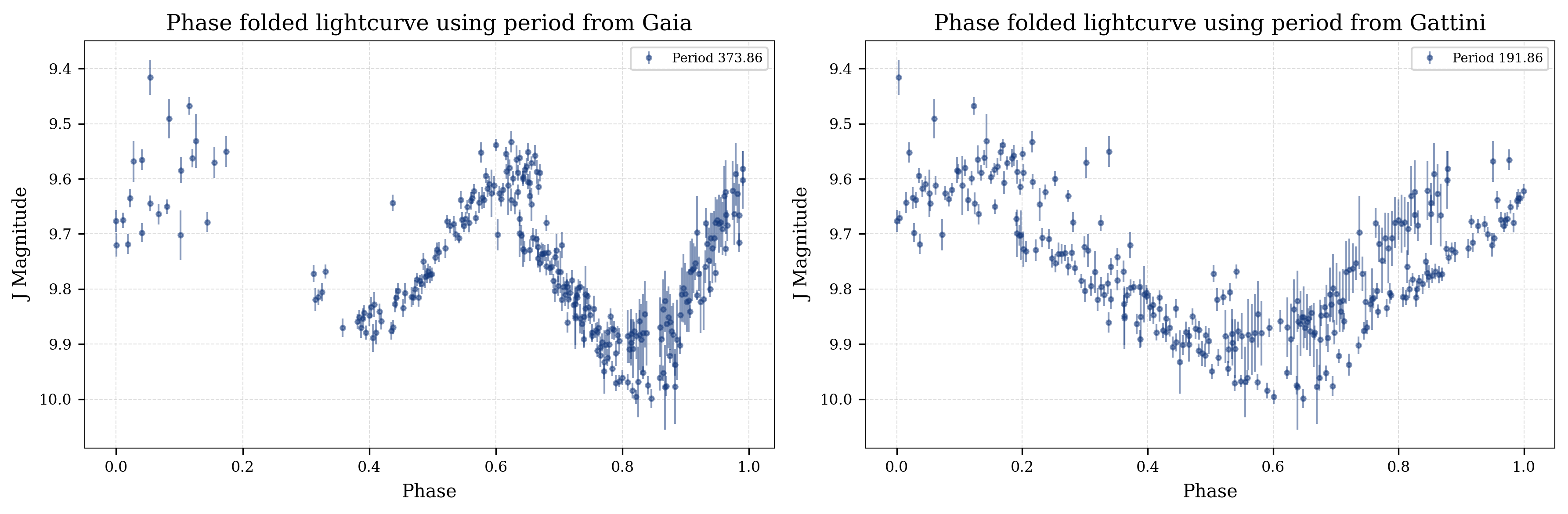}
\caption{Example of an LPV where period estimated from Gattini lightcurve is more accurate than the one calculated from Gaia data
\label{fig:gattini_right}}
\end{figure*}

The second catalog of long-period variables \citep{Lebzelter2023} from Gaia data release 3 is the largest database of LPVs to date, with 1,720,588 entries spanning 34 months of Gaia data, with 391,914 LPVs having period information. We validate our catalog by comparing it with 150,195 LPVs from Gaia, above a declination of {$-28^\circ$}, that have calculated periods. We query the Gaia archive to retrieve the positions, periods, colors, and distances (whenever available) for these LPVs. We crossmatch the Gattini LPVs with Gaia keeping a maximum offset of 6 arcsecs (corresponding to 2 PGIR pixels), to obtain 37,661 sources. Figure \ref{fig:gaia_period_xmatch} shows the comparison between periods obtained from Gattini and Gaia. We observe a strong positive correlation with some scatter around the line of equal periods. We also observe a branch of correlated periods, where the Gaia period is approximately twice as much as the Gattini period, like the example shown in Figure \ref{fig:gattini_right}. Visual inspection of these sources shows that Gaia overestimates the period of these variables, likely due to sparser coverage than PGIR. 

Additionally, we also crossmatch our catalog with SIMBAD to understand the classification accuracy of the machine learning model. The SIMBAD website provides a collection of almost all variables previously found in the Catalina, OGLE, Gaia, and ASAS surveys; among others. We consider our LPV classification to be correct for a star if its previously assigned category from other surveys falls into one of 'LongPeriodV*', 'AGB*', 'C*', 'Mira', 'LongPeriodV*\_Candidate', 'C*\_Candidate', 'Variable*', 'Star', 'OH/IR*', 'MidIR', 'NearIR', 'Mira\_Candidate' or 'RGB*'. Using a maximum crossmatch radius of 6 arcsecs, we see that 73,346 (45.9\% of the total LPVs) objects in the PGIR catalog are new. Of the remaining 84,134 objects, 1,867 fall into categories other than the ones mentioned before, indicating a high purity (97.78\%) of the PGIR catalog. The populations of the different Simbad categories are listed in Table \ref{tab:simbad}.


\begin{table}[ht]
    \centering
    \begin{threeparttable}
    \begin{tabular}{cl}
    \toprule
    Category & Counts\\
    \midrule
    New & 73,346 \\
    LongPeriodV*\_Candidate & 40,195 \\
    LongPeriodV* & 22,386 \\
    Mira & 18,020 \\
    Star & 1,845 \\
    C* & 805 \\
    Variable* & 409 \\
    AGB* & 270 \\
    OH/IR* & 220 \\
    MidIR/NearIR & 127 \\
    C*\_Candidate & 122 \\
    RGB* & 82 \\
    Mira\_Candidate & 2 \\
    Others & 1,867\\
    \bottomrule
    \end{tabular}
    \begin{tablenotes}
        \item \small{
    Objects with confirmed sub-types are counted into their respective sub-types (e.g. Mira) and excluded from the broader class (e.g. LongPeriodV*)}
    \end{tablenotes}
    \end{threeparttable}
    \caption{Simbad types of objects in the PGIR LPV catalog}
    \label{tab:simbad}
\end{table}

%% file: conclusion.tex
We present a catalog of 159,696 Long Period Variables from the PGIR survey created using a machine-learning classifier. We assemble a training set of 4300 objects spanning LPVs showing clean sinusoidal evolution in their lightcurves, ``Type II LPVs" showing non-sinusoidal but periodic lightcurves, non-LPVs comprised of R Cor Bor stars, Young Stellar Objects and other classes that display long timescale erratic variability, and non-variables. We train a gradient-boosted decision tree classifier on an artificially resampled dataset using Adaptive synthetic upsampling and allKNN downsampling to class-balance the training set, achieving a true positive rate of 94.89\% for the combined set of LPVs and Type II LPVs and an overall weighted g-mean score of 0.95, indicating high accuracy.

Following a feature extraction process for 35 million lightcurves from the PGIR survey at a rate of ~0.1 seconds per lightcurve, we generate the LPV catalog using classifier predictions. We validate the catalog by crossmatching it with the Gaia catalog of LPVs from the third data release \citep{Lebzelter2023} and the Simbad interface to ascertain purity and completeness. We find that 73,346 variables are newly discovered, which is 45.9\% of the total catalog size, and find that 1,867 objects in the catalog fall out of desired Simbad classification categories, indicating a purity of \til97.8\%. We also find a good correlation in periods between LPVs from PGIR and Gaia. We use the PGIR LPV catalog to examine the period-amplitude relationship for Miras in the J band and verify the bimodality in period distributions of Miras found in other studies \citep{,Matsunaga05,Matsunaga09,Glass01,Nikzat22}. The methods developed in this paper will be useful for more sensitive searches for periodic variables with  the Vera Rubin Observatory \citep{Ivezic2019}.

%% file: acknowledgement.tex
Palomar Gattini-IR (PGIR) is generously funded by Caltech, Australian National University, the Mt Cuba Foundation, the Heising Simons Foundation, the Binational Science Foundation. PGIR is a collaborative project among Caltech, Australian National University, University of New South Wales, Columbia University and the Weizmann Institute of Science. MMK acknowledges generous support from the David and Lucille Packard Foundation. MMK acknowledges the US-Israel Bi-national Science Foundation Grant 2016227. MMK acknowledges the Heising-Simons foundation for support via a Scialog fellowship of the Research Corporation. MMK and AMM acknowledge the Mt Cuba foundation. JS is supported by an Australian Government Research Training Program (RTP) Scholarship. AS acknowledges support from the Caltech Summer Undergraduate Research Fellowship (SURF) funded by the NSF Astronomy and Astrophysics grant number 2206730. KD was supported by NASA through the NASA Hubble Fellowship grant \#HST-HF2-51477.001 awarded by the Space Telescope Science Institute, which is operated by the Association of Universities for Research in Astronomy, Inc., for NASA, under contract NAS5-26555. RS acknowledges grant number 12073029 from the National Natural Science Foundation of China (NSFC). 